\documentclass[aps,showpacs,superscriptaddress,twocolumn]{revtex4}
\usepackage{graphicx}
\begin{document}

\title{Global climate models violate scaling of the observed atmospheric 
variability}

\author{R.~B.~Govindan}
\email{govindan@yafit.ph.biu.ac.il}
\affiliation{Minerva Center and Department of Physics, 
Bar-Ilan University, Ramat-Gan 52900, Israel}
\affiliation{Institut f\"ur Theoretische Physik~III, 
Justus-Liebig-Universit\"at Giessen,
Heinrich-Buff-Ring 16, 35392 Giessen, Germany}

\author{\surname{Dmitry} Vyushin}
\affiliation{Minerva Center and Department of Physics, 
Bar-Ilan University, Ramat-Gan 52900, Israel}
\affiliation{Institut f\"ur Theoretische Physik~III, 
Justus-Liebig-Universit\"at Giessen,
Heinrich-Buff-Ring 16, 35392 Giessen, Germany}

\author{\surname{Armin} Bunde}
\email{Armin.Bunde@theo.physik.uni-giessen.de}
\affiliation{Institut f\"ur Theoretische Physik~III, 
Justus-Liebig-Universit\"at Giessen,
Heinrich-Buff-Ring 16, 35392 Giessen, Germany}

\author{\surname{Stephen} Brenner}
\affiliation{Department of Geography, Bar-Ilan University, Ramat-Gan 52900, Israel}

\author{\surname{Shlomo} Havlin}
\affiliation{Minerva Center and Department of Physics, 
Bar-Ilan University, Ramat-Gan 52900, Israel}

\author{\surname{Hans-Joachim} Schellnhuber} 
\affiliation{Potsdam Institute for Climate Impact Research, 
D-14412 Potsdam, Germany}

\date{\today}

\begin{abstract}
We test the scaling performance of seven leading global climate 
models by using detrended fluctuation analysis. We analyse 
temperature records of six representative sites around the globe 
simulated by the models, for two different scenarios: (i)~with 
greenhouse gas forcing only and (ii)~with greenhouse gas plus 
aerosol forcing. We find that the simulated records for both
scenarios fail to reproduce the universal scaling behavior of the 
observed records, and display wide performance differences. 
The deviations from the scaling behavior are more pronounced in 
the first scenario, where also the trends are clearly 
overestimated. 
\end{abstract}
\maketitle

\noindent
Confidence in the simulation and prediction skills of global climate models (coupled atmosphere-ocean general 
circulation models~\cite{1,Has} AOGCMs) is a crucial precondition for formulating climate protection policies.
The models provide numerical solutions of the Navier--Stokes 
equations devised for simulating meso-scale to large-scale atmospheric and oceanic 
dynamics. In addition to the explicitly resolved scales of motions, the models 
also contain parameterization schemes representing the so-called subgrid-scale 
processes, such as radiative transfer, turbulent mixing, boundary layer processes, 
cumulus convection, precipitation, and gravity wave drag. A radiative transfer scheme, 
for example, is necessary for simulating the role of various greenhouse gases such 
as CO$_2$ and the effect of aerosol particles.   
The differences among the models usually lie in the selection of the 
numerical methods employed, the choice of the spatial resolution~\cite{FN1}, and the 
subgrid-scale parameters. 

Two scenarios (apart from a control run with fixed CO$_2$ content) have been studied by the models, 
and the results are available from the IPCC Data Distribution Center~\cite{26}.
In the first scenario, one considers only the effect of greenhouse gas forcing. 
The amount of greenhouse gases is taken from the observations  until 1990 and then 
increased at a rate of 1\%
per year. In the second scenario, also the effect of aerosols (mainly sulphates)
in the atmosphere is taken into account. Only direct sulphate
forcing  is considered;  until 1990, the sulphate concentrations are taken from
historical measurements, and  are increased linearly afterwards. The effect of sulphates is to mitigate and 
partially offset the greenhouse gas warming.
Although this scenario represents an important step
towards comprehensive climate simulation, it introduces
new uncertainties --- regarding the distributions of natural and
anthropogenic aerosols and, in particular, regarding indirect effects
on the radiation balance through cloud cover modification, 
\textit{etc.}~\cite{WGI}.

All of the models are capable, to varying extents, to reproduce the
current mean state of the atmosphere~\cite{fn2}.
The models have been validated by comparing to historical data and by intercomparison
of the models~\cite{3,9}. The efforts have been restricted to traditional time series 
analysis which generally assumes that the statistical 
properties of a signal remain the same throughout  the entire  period. This 
assumption of stationarity, however, is certainly not valid for climate records 
due to imposed effects such as global or urban warming. 

In our evaluation of the models, we apply
detrended fluctuation analysis (DFA)~\cite{11,13} 
which can distinguish between trends and correlations and thus reveal 
trends as well as long term correlations very often masked by nonstationarities. 
Recently, Koscielny-Bunde \textit{et al.}~\cite{2,14} applied DFA and wavelet 
techniques (see, \textit{e.~g.}~\cite{12}) to investigate 
temporal correlations in the atmospheric variability. Considering maximum daily 
temperature records of various stations around the globe, 
they analyzed the temperature variations from their average values and found that 
the persistence, characterized by the correlation $C(s)$ between temperature 
variations separated by $s$ days, decays with a power law, 
\begin{equation}\label{eq1}
C(s) \sim s^{-\gamma},
\end{equation}                                                           
with roughly the same correlation exponent  $\gamma \cong 0.7$ for all stations considered. 
The range of this persistence law exceeds ten years, and there is no evidence 
for a breakdown of the law at even larger time scales. Indications for the long 
term persistence through spectral analysis have also been obtained~\cite{15,Talkner}. 
Since the persistence scaling law appears to be universal, \textit{i.~e.} independent of the  
location and climatic zone of the stations, we use it in the following for assessing the 
performance of the AOGCMs.

For the test, we consider monthly averages of the daily maximum temperature from  seven AOGCMs: 
GFDL-R15-a (Princeton), CSIRO-Mk2 (Melbourne), ECHAM4/OPYC3 (Hamburg), HADCM3 (Bracknell, UK), 
CGCM1 (Victoria, Canada), CCSR/NIES (Tokyo),  NCAR PCM (Boulder, USA) (see~\cite{26} for details). 
We extracted the data for six representative sites around the globe (Prague, Kasan, Seoul, 
Luling/Texas, Vancouver, and Melbourne). For each model and each scenario, we selected the temperature 
records of the 4 grid points closest to each site, and bilinearly interpolated the 
data to the location of the site. We analyze both scenarios but we focus  more on
the better established first scenario.

We analyze for each site the variations $\Delta T_i$ of the 
monthly temperatures $T_i$ from the respective monthly mean temperature $\overline{T_i}$ 
that has been obtained by averaging over all years in the record. Quantitatively, 
persistence in $\Delta T_i$ can be characterized by the (auto) correlation function, 
$
C(s) \equiv  \langle \Delta T_i \Delta T_{i+s} \rangle = 
(1/(N-s))\sum_{i=1}^{N-s} \Delta T_i \Delta T_{i+s}, 
$                                          
where $N$ is the total number of months in the record. A direct calculation of $C(s)$ 
is hindered by the level of noise present in the finite temperature series, and by 
possible nonstationarities in the data. Following Refs.~\cite{28} and~\cite{29}, 
we do not calculate $C(s)$ directly, but instead  study the fluctuations in the 
temperature ``profile'' $Y_n = \sum_{i=1}^n \Delta T_i$.
To this end, we divide the profile into (nonoverlapping) segments of length $s$ and 
determine the squared fluctuations of the profile (specified below) in each segment. 
The mean square fluctuations, averaged over all segments of length $s$, are related 
to the correlation function $C(s)$ (see below). In our test, we employ a hierarchy of methods 
that differ in the way the fluctuations are measured and possible nonstationarities are eliminated
(for a detailed description of the methods we refer to~\cite{13}) :
\begin{enumerate}
\renewcommand{\theenumi}{\roman{enumi}}
\renewcommand{\labelenumi}{(\theenumi)}
\item In the (standard) fluctuation analysis (FA), we calculate the difference of the 
profile at both ends of each segment. The square of this difference represents the 
square of the fluctuations in each segment.
\item In the ``first order detrended fluctuation analysis'' (DFA1), we determine in each 
segment the best linear fit of the profile. The standard deviation of the profile 
from this straight line represents the square of the fluctuations in each segment.
\item More generally, in the ``$n$th order DFA'' (DFA$n$), we determine in each segment the 
best $n$th-order polynomial fit of the profile. Again, the standard deviation of the 
profile from these polynomials represents the square of the fluctuations in each segment.
\end{enumerate}
The fluctuation function $F(s)$ is the root mean square of the fluctuations in all 
segments. For the relevant case of long-term power-law correlations given by Eq.~\ref{eq1}, 
with $0<\gamma<1$, the fluctuation function $F(s)$ increases 
according to a power law~\cite{30},
\begin{displaymath}
F(s) \sim s^\alpha, \qquad\alpha = 1-\frac{\gamma}{2}. 
\end{displaymath} 
For uncorrelated data (as well as for short-range correlations represented by $\gamma \geq 1$ 
or exponentially decaying correlation functions), we have $\alpha = \frac{1}{2}$.  For long term 
correlations we have $\alpha > \frac{1}{2}$.
\begin{figure}[h]   
\includegraphics[width=3.2in]{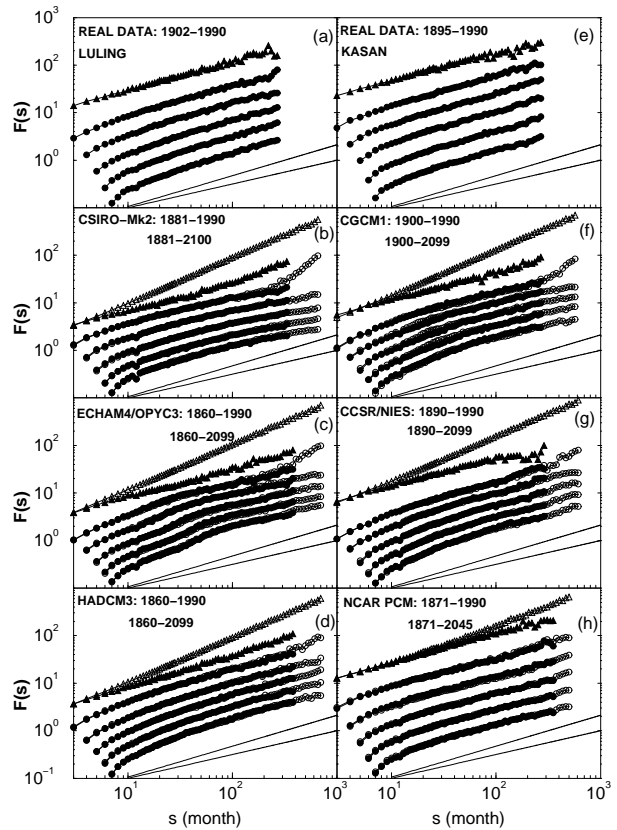} 
\caption 
{Results of FA and DFA for the monthly averages of the daily maximum temperature of Luling (a--d) 
and Kasan (e--h). In each panel, the curves from top to bottom represent the fluctuation function $F(s)$ 
obtained from FA ($\triangle$) and DFA1--DFA5 (o), respectively. Full symbols represent past data and 
open symbols are for the entire simulation period. The scale of $F(s)$ is arbitrary. The maximum value 
of $s$ is $1/4$ of the length of the considered record. The two 
lines shown at the bottom are theoretical lines with slope 0.65 (upper line) and 0.5 
(lower line). Note that FA (which does not remove trends) overestimates the fluctuation 
exponent as can be seen when comparing to DFA. As seen in the figures, the differences 
in the exponents obtained by DFA3 and higher orders of DFA are negligible, which means 
that in DFA3 all trends are removed from the data. For this reason, when focusing on the 
correlation exponent, it is sufficient to use DFA3 (as done in Figure~\ref{fig2}).   
}
\label{fig1}
\end{figure}

By definition, the FA does not eliminate trends, similar to the Hurst method and the conventional 
spectral analysis~\cite{31}. In contrast, DFAn eliminates polynomial trends of order $n-1$ 
in the original data. Thus, from the comparison of the various fluctuation functions $F(s)$ 
obtained by these methods, we can learn both about long term correlations and types of 
trends, which cannot be achieved by the conventional spectral analysis.
\begin{figure}[h]   
\includegraphics[width=3.2in]{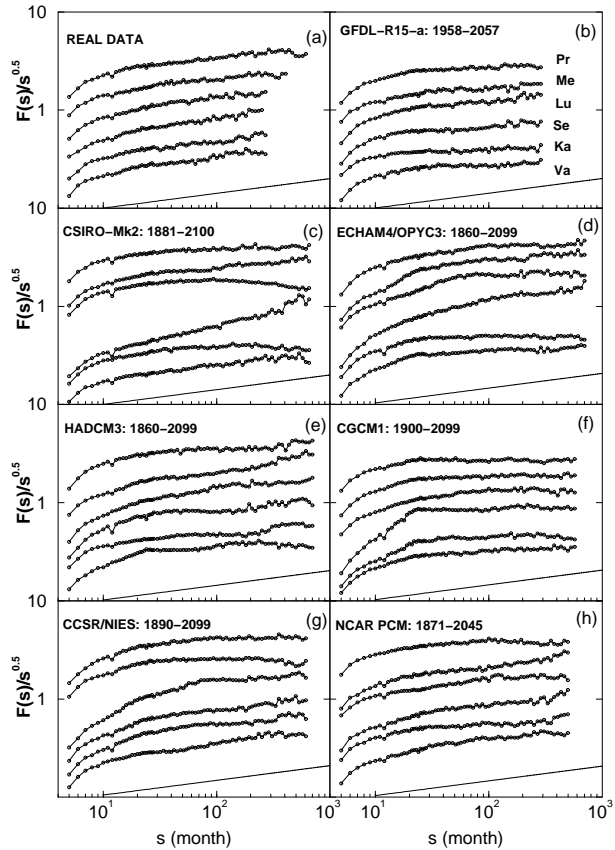} 
\caption{ 
Results of DFA3 applied to the monthly averages of the daily maximum temperature of (a) real data and (b--h) simulated 
data (for scenario~(i)) at the respective geographical positions of Prague (Pr), Melbourne (Me), Luling (Lu), 
Seoul (Se), Kasan (Ka) and Vancouver (Va), for the seven AOGCMs considered. In each panel, 
curves from top to bottom represent the result for the six sites. For better evaluation of 
the models, $F(s)$ is divided by $s^{0.5}$. The scale of $F(s)$ is arbitrary. The theoretical line 
at the bottom has a slope of 0.15.
}
\label{fig2}
\end{figure}
For testing the performance of the models we plot, in a double logarithmic presentation, 
$F(s)$ versus $s$ for FA and DFA1--DFA5 for each of the six sites and compare the curves 
with those obtained from the real data. Figure~\ref{fig1} shows representative results 
of the fluctuation functions for Kasan (Russia) and Luling (Texas) for 
both the real data and the data from six of the climate models with scenario~(i) (three models for each 
location). The data from the models 
are taken (a) until 1990 (full symbols), which is the simulation period corresponding to the 
observed data and (b) for the entire simulation period including future data (open symbols). 

Every panel shows, from top to bottom, the fluctuation functions $F(s)$ obtained from FA 
and DFA1--5. We also have drawn two straight lines with slopes 0.5 (corresponding to uncorrelated 
data) and 0.65, corresponding to the correlation exponent obtained for the real data. 
Figure~\ref{fig1} shows that for the real data, for both Kasan and Luling, \textit{all} the $F(s)$ curves are 
parallel straight 
lines with slope close to $\alpha=0.65$, beyond 1 year. This indicates (i) the absence of trends 
 and (ii) the existence of long term power law correlations consistent with 
earlier findings~\cite{2}. 

The simulated records show a quite different behavior. For Luling,
 CSIRO-Mk2 and ECHAM4/OPYC3 yield FA curves that 
are not parallel to the DFA curves, having a larger asymptotic slope, while the DFA curves show 
a crossover towards uncorrelated behavior ($\alpha=0.5$) after roughly 2 years. This indicates 
(i) an overestimation of the trends and (ii) the loss of long term correlations in the models. 
The HADCM3 model performs slightly better, with DFA curves approaching a slope of $\alpha=0.62$ 
at long times. However, when compared to real data, the FA curve bends slightly upwards at long 
times (for the past data), revealing also
an overestimation of the trend. For Kasan, the CGCM1 model yields uncorrelated behavior
at long times. CCSR/NIES and NCAR PCM show long term persistence, with an exponent $\alpha$
slightly below 0.6 in both cases. In all cases, the FA curves are straight lines, with slightly
larger exponents than the DFA curves for CGCM1 and NCAR PCM. This again points to an
overestimation of the trends by the models.
\vskip 2mm
\begin{figure}[h]
\includegraphics[width=3.2in]{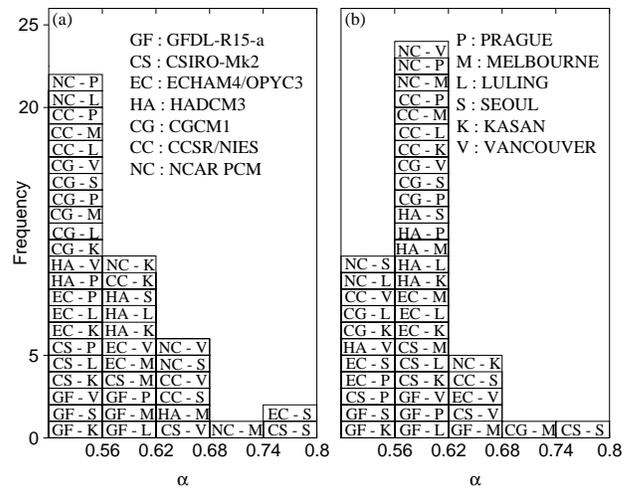} 
\caption{ 
Histogram of the fluctuation exponent ($\alpha$) values obtained for the six sites simulated by 
the seven AOGCMs (a)~scenario~(i) and (b)~scenario~(ii) for the entire records. 
The entries in each box represent `Model -- Site'.
}
\label{fig3}
\end{figure}
For reviewing the scaling performance of the models we concentrate now on the results
of DFA3, since DFA3, DFA4 and DFA5 yielded the same scaling exponents for all cases.
We find that the fluctuation function in DFA3--5 remained unchanged 
while FA is changed dramatically when including also future data (see Figure~\ref{fig1}). This feature 
shows some internal consistency of the models,
since past and future differ mainly in the amount of trends created by greenhouse gases 
(as shown by the FA curves),
and trends are well eliminated by higher order DFA~\cite{FN3}. Therefore, we use the 
entire data for a more accurate estimation of scaling exponents.
 
Figure~\ref{fig2} shows the results for the fluctuation function obtained from DFA3 for all 
available real and model data (all seven AOGCMs with scenario~(i)) at the six sites considered, 
for the entire simulation period.
 To facilitate the evaluation of the models, we have divided $F(s)$ by $s^{\frac{1}{2}}$. 
A plateau now indicates loss of long term correlations. The straight line in each panel has 
a slope of 0.15, corresponding to the universal exponent in the original curves. While the 
real data yield, for long times, parallel lines with a slope of 0.15 for all sites in 
agreement with the earlier findings, the virtual data display wide performance differences 
and fail to reproduce the universal features of the benchmark time series. Direct inspection 
of the figure shows that about half of the model curves are very close to a plateau, yielding 
uncorrelated behavior above approximately 2 years. 

The actual long term exponents $\alpha$ for the greenhouse gas only
scenario obtained by the seven models for the six sites are summarized in a histogram in Figure~\ref{fig3}a. 
The histogram shows a pronounced maximum at $\alpha=0.5$. For best performance, 
all models should have exponents $\alpha$  close to 0.65, corresponding to a peak of 
height 42 in the window between 0.62 and 0.68. Actually more than half of the exponents
are close to 0.5, while only 6 exponents are in the proper window between 0.62 and 0.68. 

Figure~\ref{fig3}b shows the histogram for scenario~(ii), where in addition to the greenhouse gas forcing, 
also the effects of aerosols are taken into 
account. For this case, there is a pronounced maximum in the $\alpha$ window between 0.56 and 0.62 
(more than half of the exponents are in this window), while only 5
exponents are in the proper range between 0.62 and 0.68. This shows that 
although the second scenario is also far from reproducing the scaling behavior of the real data, its
overall performance is better than the performance of the first scenario.

To summarize, we have presented evidence that AOGCMs fail to reproduce the universal scaling behavior 
observed in the real temperature records. Moreover, the models display wide differences in scaling for
different sites. When comparing the two scenarios, our results suggest that the second scenario
(CO$_2$ plus aerosols) exhibits better performance regarding the values of the scaling 
exponents as well as the trends. The effect of aerosols not only decreases
the trends but also modifies the fluctuations, to more closely resemble the real data. 
This confirms in a way independent of the evaluations made so far~\cite{WGI}
that the incorporation of aerosols is necessary to approach reality. 

It is possible that the lack of long-term persistence is due to the fact that
certain forcings like volcanic eruptions or solar fluctuations have not been
incorporated in the models. However, we cannot rule out that
\textit{systematic} model deficiencies (such as the use of equivalent CO$_2$
forcing to account for all other greenhouse gases or inaccurate spatial and
temporal distributions of sulphate emissions) prevent the AOGCMs from
correctly simulating the natural variability of the atmosphere.

\begin{acknowledgments}
This work has been supported by the Deutsche Forschungsgemeinschaft and the Israel Science 
Foundation. We like to thank Peter Cox, Hartmut Grassl and John Mitchell for valuable 
comments on the manuscript.
\end{acknowledgments}


\begin{thebibliography}{}
\bibitem{FN1}
A typical climate module within the AOGCM machinery will have a grid spacing of 
300--500 km and 10--20 vertical layers as compared to a weather forecasting model with 
a grid spacing of 100 km or less and 30--40 layers.
\bibitem{FN3}
The large differences between FA and DFA reemphasize that FA (like Hurst method)
can not be used by itself for evaluating the scaling exponent.
\end{thebibliography}
\end{document}